# Superalgebraic structure of Lorentz transformations

**Monakhov V.V.**
Saint Petersburg State University, Saint Petersburg, Russia;

*E-mail:* v.v.monahov@spbu.ru;

Modern relativistic theory of the second quantization of fermion and boson fields is based on the use of the mathematical apparatus of C*-algebras and Lie superalgebras. In this case, for fermions, the Lorentz transformations are considered as Bogolyubov transformations of creation and annihilation operators. However, in this approach one can not obtain an explicit form of the Dirac gamma-matrices.

The mathematical apparatus of the superalgebraic representation of the algebra of the second quantization of spinors is developed in the article. It is based on the use of density in the impulse space of Grassmann variables and their derivatives. It is shown that the Dirac matrices and the Lorentz transformation generators can be expressed in terms of such densities. A superalgebraic form of the Dirac equation and the vacuum state vector are constructed. It is shown that in the superalgebraic form of the complex Clifford algebra the generators corresponding to the Dirac gamma matrices are not equivalent. Clifford vector corresponding to diagonal matrix $\gamma^0$ annihilates the vacuum, and the remaining ones give nonzero values. This means that there is asymmetric direction corresponding to the time axis

<u>Key words:</u> spinors, Lie superalgebra, Lorentz transformations, Dirac matrices, Dirac equation, secondary quantization, creation-annihilation operators, Clifford algebra.

**DOI:**

Modern relativistic theory of the second quantization of fermion and boson fields is based on the use of the mathematical apparatus of C*-algebras [1,2] and Lie superalgebras [3]. In this case, for fermions, the Lorentz transformations are considered as Bogolyubov transformations of creation and annihilation operators [3]. However, in this approach one can not obtain an explicit form of the Dirac gamma-matrices.

In [4, 5] the author proposed a superalgebraic representation of spinors and Dirac matrices, which makes it possible to give a unified algebraic interpretation to the spinors and Dirac matrices. However, in the second quantization method, two terms correspond to birth operators, and two terms correspond to the annihilation operators, and they are differently transformed under Lorentz transformations. In [4, 5], this distinction was an external condition that does not follow from the algebraic nature of the spinors.

In the framework of the superalgebraic approach, when using Grassmann variables $\theta^\alpha$ and derivatives with respect to them $\frac{\partial}{\partial \theta^\alpha}$, at rest, the birth operators are naturally considered operators of multiplication by Grassmann variables $\theta^\alpha$, and the annihilation operators are derivatives with respect to them $\frac{\partial}{\partial \theta^\alpha}$.

Let us consider infinitesimal rotation operators $e^{iG^{\mu\nu} d\omega_{\mu\nu}}$ with generators $G^{\mu\nu}$ that take one spinor state vector to another spinor state vector. With them, the state vector $\Phi$ is transformed as $\Phi' = e^{iG^{\mu\nu} d\omega_{\mu\nu}} \Phi = (1 + iG^{\mu\nu} d\omega_{\mu\nu})\Phi$.

Denote $G^{\mu\nu} d\omega_{\mu\nu}$ by $dG$. The law of transformation of an arbitrary operator $T$ acting on the spinor state vector $\Phi$, and in doing this, again transforming it into the spinor state vector, neglecting infinitesimal higher order, will be

$$T' = e^{idG} T e^{-idG} = T + i[dG, T] = (1 + i[dG, *])T,$$

where * denotes the place to which you want to substitute the operator on which the transformation operates. If it consists of the product of several operators, then each of them is transformed according to the same law:

$$(1+[i\,dG,*])T_1T_2...T_m\Phi = (T_1+[i\,dG,T_1])(T_2+[i\,dG,T_2])...(T_m+[i\,dG,T_m])(1+i\,dG)\Phi. \quad (1)$$

It follows from (1) that if there are generators of Lorentz transformations for state vectors, operators that preserve the spinors by spinors are transformed by means of commutators with these generators. Since in the second quantization the solutions of the Dirac equation are treated as operators, they must also be transformed by such commutators. The Lorentz transformation causes a simultaneous change in the basis for all factors, acting on each of them. We denote such operators $[A,*]$ by $\hat{A}$. An operator $\hat{A} = [A,*]$ is simply a Lie derivative (commutator) acting on operators.

The formula for finite angles of rotation is obtained by integrating infinitesimal transformations:

$$T' = e^{i\int dG} T e^{-i\int dG} = e^{iG}T_1 e^{-iG} e^{iG}T_2 e^{-iG}...e^{iG}T_m e^{-iG} = T_1'T_2'...T_m', \quad (2)$$

where $G = \int dG$ and $T_r' = e^{iG} T_r e^{-iG}$ for each $r$. Note that (2) can be written in another form:

$$T' = e^{i\int [dG,*]} T = (e^{i\int [dG,*]}T_1)(e^{i\int [dG,*]}T_2)...(e^{i\int [dG,*]}T_m) = T_1'T_2'...T_m'.$$

In this case, the parentheses limit the scope of the operators $[dG,*]$. Denoting $\hat{G} = [G,*] = \int [dG,*]$, we obtain

$$T' = e^{i\hat{G}} T = (e^{i\hat{G}}T_1)(e^{i\hat{G}}T_2)...(e^{i\hat{G}}T_m) = T_1'T_2'...T_m'.$$

Variables $\theta$ and derivatives $\frac{\partial}{\partial \theta}$ are operators, so for infinitesimal transformations in accordance with (1) they must be transformed as

$$\theta' = (1+[iG^{\mu\nu} d\omega_{\mu\nu},*])\theta$$

and

$$(\frac{\partial}{\partial \theta})' = (1+[iG^{\mu\nu} d\omega_{\mu\nu},*])\frac{\partial}{\partial \theta}.$$

It is easy to see that in the secondary quantization formalism operators $[\theta^\alpha \frac{\partial}{\partial \theta^\beta},*]$ are analogous to $\theta^\alpha \frac{\partial}{\partial \theta^\beta}$ used in [4, 5]. They correspond to matrices mixing $\theta^\alpha$ and $\theta^\beta$, as well $\frac{\partial}{\partial \theta^\alpha}$ and $\frac{\partial}{\partial \theta^\beta}$. From these operators it is possible to construct matrices corresponding to spatial rotations.

Boosts should mix the "big" and "small" components of the spinor, and therefore they must mix $\frac{\partial}{\partial \theta^\alpha}$ and $\theta^\tau$. These are the operator $[\frac{\partial}{\partial \theta^\alpha}\frac{\partial}{\partial \theta^\tau},*]$ that transforms $\theta^\tau$ to $\frac{\partial}{\partial \theta^\alpha}$ and $\theta^\alpha$ to $-\frac{\partial}{\partial \theta^\tau}$, and the operator $[\theta^\alpha\theta^\tau,*]$ that transforms $\frac{\partial}{\partial \theta^\alpha}$ to $-\theta^\tau$ and $\frac{\partial}{\partial \theta^\tau}$ to $\theta^\alpha$.

So far we have considered Grassmann variables $\theta^\alpha$ and their derivatives $\frac{\partial}{\partial \theta^\alpha}$ as global quantities that do not depend on coordinates and impulses. With this interpretation, a problem arises, since the standard commutation relations for the creation and annihilation operators

$$\{b_\alpha(p), b_\beta^+(p')\} = \delta^\alpha_\beta \delta^3(p-p'),$$
$$\{d_\tau(p), d_\rho^+(p')\} = \delta^\tau_\rho \delta^3(p-p'), \quad (3)$$
$$\{b_\alpha(p), d_\tau(p')\} = \{b_\alpha(p), d_\tau^+(p')\} = \{b_\alpha^+(p), d_\tau(p')\} = \{b_\alpha^+(p), d_\tau^+(p')\} = 0$$

contain a three-dimensional delta function $\delta^3(p-p') = \delta(p_1-p_1')\delta(p_2-p_2')\delta(p_3-p_3')$, but the anticommutator of Grassmann variables and derivatives with respect to them does not allow us to obtain delta functions. Therefore, we introduce in the momentum space Grassmann variables $\theta^a(p)$ that depend on the impulse, and local derivatives with respect to them $\frac{\partial}{\partial \theta^b(p)}$, which satisfy conditions

$$\{\frac{\partial}{\partial \theta^b(p)}, \theta^a(p')\} = \delta_b^a \delta^3(p-p'),$$

$$\{\frac{\partial}{\partial \theta^a(p)}, \frac{\partial}{\partial \theta^b(p')}\} = \{\theta^a(p), \theta^b(p')\} = 0.$$

They can be regarded as densities in impulse space. Consider the operator

$$[M,*] = \int d^3p \, m_a^{\ b} [\theta^a(p) \frac{\partial}{\partial \theta^b(p)}, *].$$

When it acts on an element of the algebra of spinors

$$\chi = \int d^3p \, \chi_c(p) \theta^c(p),$$

where $\chi_c(p)$ are numerical coefficients, we obtain

$$[M,*] \int d^3p \, \chi_c(p) \theta^c(p) = \int d^3p \, m_a^{\ b} \chi_b(p) \theta^a(p),$$

but when we act on an element

$$\varphi = \int d^3p \, \varphi^a(p) \frac{\partial}{\partial \theta^a(p)},$$

where $\varphi^a(p)$ are numerical coefficients, we obtain

$$[M,*]\varphi = -\int d^3p \, \varphi^a(p) m_a^{\ b}(p) \frac{\partial}{\partial \theta^b(p)}.$$

Similarly, $\theta^a(p)$ and $\frac{\partial}{\partial \theta^b(p)}$ are mixing by operators

$$[A,*] = \int d^3p \, a_{ab}(p) [\theta^a(p) \theta^b(p), *]$$

and

$$[B,*] = \int d^3p \, b^{ab}(p) [\frac{\partial}{\partial \theta^b(p)} \frac{\partial}{\partial \theta^a(p)}, *].$$

We denote the matrix column-spinor, in which the upper element is 1, and the remaining elements are equal to zero, as $u_1$, a column in which 1 element is on the second line, and the remaining 0 as $u_2$, etc. We will make a mapping of the matrix spinor-columns $u_k$, $k=1,2,3,4$ to the operators $\theta^a(p)$ and $\frac{\partial}{\partial \theta^a(p)}$ corresponding to the given spatial impulse $p$:

$$u_1 \cong \frac{\partial}{\partial \theta^1(p)}, u_2 \cong \frac{\partial}{\partial \theta^2(p)}, u_3 \cong \theta^3(p), u_4 \cong \theta^4(p).$$

This is some simplification, since it is necessary to consider the sum of the operators corresponding to all impulses. But the transformations of the elements corresponding to different impulses occur independently, so this simplification is very convenient.

A mapping is obtained between the operators $[M,*]$, $[A,*]$, $[B,*]$ and the elements of the matrix space generated by Dirac gamma matrices. Each such operator can be associated with the same matrix, regardless of the impulse values corresponding to the spinor elements to which this operator acts.

The Hermitian conjugation is analogous to the Hermitian conjugation of superalgebraic spinors, described by the author in [4,5]: when conjugated, the numerical factors are replaced by complex conjugates, $\theta^a(p)$ are replaced by $\dfrac{\partial}{\partial \theta^a(p)}$, and $\dfrac{\partial}{\partial \theta^a(p)}$ are replaced by $\theta^a(p)$. In this way

$$(\theta^a(p))^+ = \frac{\partial}{\partial \theta^a(p)},$$

$$\left(\frac{\partial}{\partial \theta^a(p)}\right)^+ = \theta^a(p).$$

It is obvious that

$$\{\theta^a(p'), (\theta^b(p))^+\} = \{\theta^a(p'), \frac{\partial}{\partial \theta^b(p)}\} = \delta^a_b \delta^3(p - p').$$

Thus, if there are spinors

$$\Psi(p') = \psi_\alpha(p') \frac{\partial}{\partial \theta^\alpha(p')} + \psi_\tau(p') \theta^\tau(p')$$

and

$$\Phi(p) = \varphi_\beta(p) \frac{\partial}{\partial \theta^\beta(p)} + \varphi_\rho(p) \theta^\rho(p),$$

then

$$\Psi^+(p') = \psi_\alpha^*(p') \theta^\alpha(p') + \psi_\tau^*(p') \frac{\partial}{\partial \theta^\tau(p')},$$

and

$$\{\Psi^+(p'), \Phi(p)\} = (\psi_\alpha^*(p) \varphi_\alpha(p) + \psi_\tau^*(p) \varphi_\tau(p)) \delta^3(p - p'). \quad (4)$$

It follows from (4) that for spinors with a given impulse value one can define a scalar product

$$(\Psi(p'), \Phi(p)) = \{\Psi^+(p'), \Phi(p)\}. \quad (5)$$

This allows us to define an orthogonal basis in which the independent generators are orthogonal. In this case, the independent operators of spinor fields must be orthogonal. In accordance with (5), if $\Psi(p')$ and $\Phi(p)$ are independent, then $\Psi^+(p')$ and $\Phi(p)$ must anticommute.

The presence of the mapping of matrix elements to superalgebraic operators allows us to introduce operators $\hat{\gamma}^\eta, \eta = 0,1,2,3,5$, corresponding to the Dirac matrices $\gamma^\eta$ (in the Dirac representation) by the following formulas:

$$\tilde{\gamma}^0 = \int d^3p \left( -\theta^1(p) \frac{\partial}{\partial \theta^1(p)} - \theta^2(p) \frac{\partial}{\partial \theta^2(p)} - \theta^3(p) \frac{\partial}{\partial \theta^3(p)} - \theta^4(p) \frac{\partial}{\partial \theta^4(p)} \right), \quad (6)$$

$$\tilde{\gamma}^1 = \int d^3p \left( \frac{\partial}{\partial \theta^2(p)} \frac{\partial}{\partial \theta^3(p)} + \frac{\partial}{\partial \theta^1(p)} \frac{\partial}{\partial \theta^4(p)} - \theta^3(p) \theta^2(p) - \theta^4(p) \theta^1(p) \right), \quad (7)$$

$$\tilde{\gamma}^2 = i \int d^3p \left( \frac{\partial}{\partial \theta^2(p)} \frac{\partial}{\partial \theta^3(p)} - \frac{\partial}{\partial \theta^1(p)} \frac{\partial}{\partial \theta^4(p)} + \theta^3(p) \theta^2(p) - \theta^4(p) \theta^1(p) \right), \quad (8)$$

$$\tilde{\gamma}^3 = \int d^3p \left( \frac{\partial}{\partial \theta^1(p)} \frac{\partial}{\partial \theta^3(p)} - \frac{\partial}{\partial \theta^2(p)} \frac{\partial}{\partial \theta^4(p)} - \theta^3(p) \theta^1(p) + \theta^4(p) \theta^2(p) \right), \quad (9)$$

$$\tilde{\gamma}^5 = \int d^3p \left( \frac{\partial}{\partial \theta^1(p)} \frac{\partial}{\partial \theta^3(p)} + \frac{\partial}{\partial \theta^2(p)} \frac{\partial}{\partial \theta^4(p)} + \theta^3(p) \theta^1(p) + \theta^4(p) \theta^2(p) \right), \quad (10)$$

$$\hat{\gamma}^\eta = [\tilde{\gamma}^\eta, *]. \quad (11)$$

Operators $\hat{\gamma}^\eta$ convert basis elements $\dfrac{\partial}{\partial \theta^1(p)}, \dfrac{\partial}{\partial \theta^2(p)}, \theta^3(p), \theta^4(p)$ and their linear combinations in the same way as the Dirac matrices $\gamma^\eta$ convert the columns $u_k$ and their linear combinations. Moreover, a linear combination of products of an arbitrary number of operators $\hat{\gamma}^\eta$ corresponds to a matrix from the corresponding linear combination of products of matrices $\gamma^\eta$.

The situation is completely analogous to the generators of the Lorentz transformations. In the matrix form they are given by formulas

$$\sigma^{\mu\nu} = \frac{i}{2}[\gamma^\mu, \gamma^\nu], \quad \sigma^{0k} = i\begin{pmatrix} 0 & \tau^k \\ \tau^k & 0 \end{pmatrix}, \quad \sigma^{kl} = \begin{pmatrix} \tau^m & 0 \\ 0 & \tau^m \end{pmatrix}, \tag{12}$$

where $\mu, \nu = 0,1,2,3,5$, $\tau^k$ are the Pauli matrices, and the indices $k,l,m$ are cyclic permutation of 1, 2, 3. Similarly to (6)-(10), we can define $\tilde{\sigma}^{\mu\nu}$ by the mapping of the matrix elements (12) to operators:

$$\tilde{\sigma}^{01} = i\int d^3p \left( \frac{\partial}{\partial \theta^2(p)} \frac{\partial}{\partial \theta^3(p)} + \frac{\partial}{\partial \theta^1(p)} \frac{\partial}{\partial \theta^4(p)} + \theta^3(p)\theta^2(p) + \theta^4(p)\theta^1(p) \right),$$

and so on. The operators $\tilde{\sigma}^{\mu\nu}$ satisfy equality

$$\tilde{\sigma}^{\mu\nu} = \frac{i}{2}[\tilde{\gamma}^\mu, \tilde{\gamma}^\nu].$$

We define the operators that are generators of the Lorentz transformations

$$\hat{\sigma}^{\mu\nu} = \frac{i}{2}(\hat{\gamma}^\mu \hat{\gamma}^\nu - \hat{\gamma}^\nu \hat{\gamma}^\mu).$$

It is easy to verify that

$$\hat{\sigma}^{\mu\nu} = [\tilde{\sigma}^{\mu\nu}, *]. \tag{13}$$

Thus, operators $\hat{\sigma}^{\mu\nu}$ can match matrices in exactly the same way as we matched $\hat{\gamma}^\mu$ the matrix operators $\gamma^\mu$.

For an infinitesimal Lorentz transformation with generator $dG = \frac{1}{4}\tilde{\sigma}^{\mu\nu}d\omega_{\mu\nu}$, for the product of the spinor operators $\Psi_1 \Psi_2 ... \Psi_m$ it follows from equation (1) that

$$\Psi'_1 \Psi'_2 ... \Psi'_m = (\Psi_1 + i[dG, \Psi_1])(\Psi_2 + i[dG, \Psi_2])...(\Psi_m + i[dG, \Psi_m]). \tag{14}$$

In accordance with (13), the operators $\tilde{\sigma}^{\mu\nu}$ generate the Lorentz transformation generators $\hat{\sigma}^{\mu\nu}$, and each of the spinor operators in (14) is transformed as follows:

$$\Psi'_k = (1 + \frac{i}{4}\hat{\sigma}^{\mu\nu}d\omega_{\mu\nu})\Psi_k. \tag{15}$$

Spinors $\Psi_k$ are a linear combination of $\frac{\partial}{\partial \theta^1(p)}, \frac{\partial}{\partial \theta^2(p)}, \theta^3(p), \theta^4(p)$, and by construction the action of operators $\hat{\sigma}^{\mu\nu}$ on them corresponds to the action of matrices $\sigma^{\mu\nu}$ on a column. Wherein

$$(i\hat{\sigma}^{01})^2 \Psi_k = \Psi_k, \tag{16}$$

$$(i\hat{\sigma}^{rl})^2 \Psi_k = -\Psi_k. \tag{17}$$

Integrating (15) with allowance for (16), we find that a boost on a finite angle $\varphi = -\omega_{01}$ in the plane $\gamma^0, \gamma^1$ corresponds a transformation $\Psi'_k = (ch\frac{\varphi}{2} - i\hat{\sigma}^{01} sh\frac{\varphi}{2})\Psi_k$.

Similarly, from (15) and (17) it follows that for spatial rotation by an angle $\alpha$ in the plane $\gamma^r, \gamma^l$ we obtain $\Psi'_k = (\cos\frac{\alpha}{2} - i\hat{\sigma}^{rl} \sin\frac{\alpha}{2})\Psi_k$.

The operator carrying out a boost in the plane $\gamma^0, \gamma^1$ can be written in the form

$$\hat{T} = e^{\hat{\gamma}^0 \hat{\gamma}^1 \varphi/2} = e^{[\tilde{\gamma}^0, *][\tilde{\gamma}^1, *]\varphi/2}.$$

As a result of the transformation we get

$$\Psi' = e^{\hat{\gamma}^0 \hat{\gamma}^1 \varphi/2} \Psi. \tag{18}$$

Similarly for spatial rotation

$$\Psi' = e^{\hat{\gamma}^r \hat{\gamma}^l \alpha/2} \Psi.$$

The Hermitian conjugation of matrices is given by the scalar product (5), but in the general case for arbitrary impulses it is necessary to consider the anticommutator of the $\Psi^+(x') = (\int d^3 p' \, \Psi(p'))^+$ and $\Phi(x) = \int d^3 p \, \Phi(p)$, that is, to specify the scalar product of the spinor operators in the form

$$(\Psi(x'), \Phi(x)) = \{\Psi^+(x'), \Phi(x)\}. \tag{19}$$

By the definition of a Hermitian adjoint operator

$$(\Psi(x'), \hat{\gamma}^\mu \Phi(x)) = ((\hat{\gamma}^\mu)^+ \Psi(x'), \Phi(x)) \tag{20}$$

for arbitrary $\Psi(x')$ and $\Phi(x)$. Therefore, to find $(\hat{\gamma}^\mu)^+$ it is necessary to find an operator $A$ such that $(A\Psi(x'), \Phi(x)) = (\Psi(x'), \hat{\gamma}^\mu \Phi(x))$ for any admissible $\Psi(x')$ and $\Phi(x)$. We will consider the expansion $\Phi(x)$ and $\Psi(x')$ with respect to impulses in the form

$$\Phi(x) = \int d^3 p \, \Phi(p) = \int d^3 p \, (\varphi_\alpha(p) \frac{\partial}{\partial \theta^\alpha(p)} e^{-ip_\mu x^\mu} + \varphi_\tau(p) \theta^\tau(p) e^{ip_\mu x^\mu}), \tag{21}$$

$$\Psi(x') = \int d^3 p' \, \Psi(p') = \int d^3 p' (\psi_\alpha(p') \frac{\partial}{\partial \theta^\alpha(p')} e^{-ip'_\mu x'^\mu} + \psi_\tau(p') \theta^\tau(p') e^{ip'_\mu x'^\mu}).$$

Consider the action of $\hat{\gamma}_0$ on these spinors:

$$\hat{\gamma}^0 \Phi(x) = \int d^3 p (\varphi_\alpha(p) \frac{\partial}{\partial \theta^\alpha(p)} e^{-ip_\mu x^\mu} - \varphi_\tau(p) \theta^\tau(p) e^{ip_\mu x^\mu}), \tag{22}$$

$$\hat{\gamma}^0 \Psi(x') = \int d^3 p' (\psi_\alpha(p') \frac{\partial}{\partial \theta^\alpha(p')} e^{-ip'_\mu x'^\mu} - \psi_\tau(p') \theta^\tau(p') e^{ip'_\mu x'^\mu}).$$

Hermitian adjoint to $\Psi(x')$ and $\hat{\gamma}^0 \Psi(x')$ spinors:

$$\Psi^+(x') = \int d^3 p' (\psi^*_\alpha(p') \theta^\alpha(p') e^{ip'_\mu x'^\mu} + \psi^*_\tau(p') \frac{\partial}{\partial \theta^\tau(p')} e^{-ip'_\mu x'^\mu}), \tag{23}$$

$$(\hat{\gamma}^0 \Psi(x'))^+ = \int d^3 p' (\psi^*_\alpha(p') \theta^\alpha(p') e^{ip'_\mu x'^\mu} - \psi^*_\tau(p') \frac{\partial}{\partial \theta^\tau(p')} e^{-ip'_\mu x'^\mu})). \tag{24}$$

From the definition (19), and also relations (23) and (22) it follows that

$$(\Psi(x'), \hat{\gamma}^0 \Phi(x)) = \{\Psi^+(x'), \hat{\gamma}^0 \Phi(x)\} =$$
$$= \int d^3 p \, (\psi^*_\alpha(p) \varphi_\alpha(p) e^{ip_\mu (x'^\mu - x^\mu)} - \psi^*_\tau(p) \varphi_\tau(p) e^{-ip_\mu (x'^\mu - x^\mu)}).$$

Similarly, it follows from (19), (24) and (21) that

$$(\hat{\gamma}^0 \Psi(x'), \Phi(x)) = \{(\hat{\gamma}^0 \Psi(x'))^+, \Phi(x)\} =$$
$$= \int d^3 p \, (\psi^*_\alpha(p) \varphi_\alpha(p) e^{ip_\mu (x'^\mu - x^\mu)} - \psi^*_\tau(p) \varphi_\tau(p) e^{-ip_\mu (x'^\mu - x^\mu)}).$$

In this way, $(\Psi(x'), \hat{\gamma}^0 \Phi(x)) = (\hat{\gamma}^0 \Psi(x'), \Phi(x))$, therefore, from the definition (20) we obtain that

$$(\hat{\gamma}^0)^+ = \hat{\gamma}^0. \tag{25}$$

Similarly we obtain

$$(\hat{\gamma}^k)^+ = -\hat{\gamma}^k, \quad k = 1, 2, 3, \tag{26}$$

$$(\hat{\gamma}^5)^+ = \hat{\gamma}^5. \tag{27}$$

For $\tilde{\gamma}^\mu$ and for the spinor operator $\Psi(x)$, the operation of a Hermitian conjugation is given in the same way as for elements of the algebra of ordinary superalgebraic spinors [4,5]. The Hermitian conjugating relation $\hat{\gamma}^\mu \Psi = \tilde{\gamma}^\mu \Psi - \Psi \tilde{\gamma}^\mu$, we obtain

$$(\hat{\gamma}^\mu \Psi)^+ = \Psi^+ (\tilde{\gamma}^\mu)^+ - (\tilde{\gamma}^\mu)^+ \Psi^+ = -[(\tilde{\gamma}^\mu)^+, *] \Psi^+ .$$

Hermitian conjugating (6)-(10) we find that

$$(\tilde{\gamma}^0)^+ = \tilde{\gamma}^0; \quad (\tilde{\gamma}^k)^+ = -\tilde{\gamma}^k, \quad k=1,2,3; \quad (\tilde{\gamma}^5)^+ = \tilde{\gamma}^5 .$$

Therefore

$$[(\tilde{\gamma}^0)^+, *] = [\tilde{\gamma}^0, *] = \hat{\gamma}^0 = (\hat{\gamma}^0)^+ ,$$
$$[(\tilde{\gamma}^k)^+, *] = -[\tilde{\gamma}^k, *] = -\hat{\gamma}^k = (\hat{\gamma}^k)^+ ,$$

and

$$[(\tilde{\gamma}^5)^+, *] = [\tilde{\gamma}^5, *] = \hat{\gamma}^5 = (\hat{\gamma}^5)^+ .$$

That is

$$[(\tilde{\gamma}^\mu)^+, *] = [\tilde{\gamma}^\mu, *]^+ = (\hat{\gamma}^\mu)^+ ,$$

whereby

$$(\hat{\gamma}^\mu \Psi)^+ = -(\hat{\gamma}^\mu)^+ \Psi^+ . \tag{28}$$

So that

$$(\hat{\gamma}^\mu \hat{\gamma}^\nu \Psi)^+ = -\hat{\gamma}^{\mu+} (\hat{\gamma}^\nu \Psi)^+ = \hat{\gamma}^{\mu+} \hat{\gamma}^{\nu+} \Psi^+ . \tag{29}$$

It follows from (29) and (27) that

$$(\hat{\gamma}^k \hat{\gamma}^l \Psi)^+ = \hat{\gamma}^{k+} \hat{\gamma}^{l+} \Psi^+ = \hat{\gamma}^k \hat{\gamma}^l \Psi^+ , \tag{30}$$
$$(\hat{\gamma}^0 \hat{\gamma}^k \Psi)^+ = \hat{\gamma}^{0+} \hat{\gamma}^{k+} \Psi^+ = -\hat{\gamma}^0 \hat{\gamma}^k \Psi^+ , \tag{31}$$

where $k, l = 1, 2, 3$.

The relations (25)-(27) are completely analogous to the matrix, but (28)-(31) essentially differs from the usual matrix rules of Hermitian conjugation. This is due to the fact that in this approach there are no right-to-left operators, and all operators operate from left to right. The Dirac equation in the superalgebraic representation of the second quantization is written in the form

$$i\hat{\gamma}^\mu \partial_\mu \Psi = m\Psi . \tag{32}$$

It is easy to show that in the Dirac adjoint spinor $\overline{\Psi} = (\hat{\gamma}^0 \Psi)^+ = -\hat{\gamma}^0 \Psi^+$ satisfies the analogous equation:

$$i\hat{\gamma}^\mu \partial_\mu \overline{\Psi} = m\overline{\Psi} . \tag{33}$$

It must be taken into account that in the superalgebraic representation the repeated Dirac conjugation changes sign before the spinor: $\overline{\overline{\Psi}} = -\Psi$.

Solutions (32) and (33) can be written in the form

$$\Psi(t, x) = \int \frac{d^3 p}{\sqrt{(2\pi)^3}} (\lambda_\alpha b_\alpha(p) e^{-ip_\mu x^\mu} + \lambda_\tau \overline{d}_\tau(p) e^{ip_\mu x^\mu}), \tag{34}$$

$$\overline{\Psi}(t, x) = \int \frac{d^3 p}{\sqrt{(2\pi)^3}} (-\lambda_\tau^* d_\tau(p) e^{-ip_\mu x^\mu} + \lambda_\alpha^* \overline{b}_\alpha(p) e^{ip_\mu x^\mu}),$$

where $\lambda_\alpha$ and $\lambda_\tau$ are constants (possibly depending on $p$), the annihilation operators $b_\alpha(p)$ and $d_\tau(p)$ are obtained by Lorentz rotations of $\frac{\partial}{\partial \theta^\alpha(0)}$ and $\frac{\partial}{\partial \theta^\tau(0)}$, the creation operators $\overline{b}_\alpha(p)$ and $\overline{d}_\tau(p)$ are Lorentz rotations of $\theta^\alpha(0)$ and $\theta^\tau(0)$. In (34), instead of the notation $d_\tau^+(p)$ used in the matrix theory and, in particular, in (3), we use $\overline{d}_\tau(p)$, since $b_\alpha(p)$ and $\overline{b}_\alpha(p)$ under the Lorentz transformations are transformed covariantly, while from (18) and (31) it follows that $b_\alpha^+(p)$ transforms noncovariantly, and therefore can not be an operator acting on spinors. The same applies to $d_\tau^+(p)$.

We divide the impulse space into infinitesimal volumes $\Delta^3 p_j$. We denote

$$B_\alpha(p_j) = \frac{1}{\Delta^3 p_j} \int_{\Delta^3 p_j} d^3 p \, b_\alpha(p), \qquad (35)$$

$$D_\tau(p_j) = \frac{1}{\Delta^3 p_j} \int_{\Delta^3 p_j} d^3 p \, d_\tau(p), \qquad (36)$$

The factor $\frac{1}{\Delta^3 p_j}$ is necessary for the $B_\alpha(p_j)$ and $D_\tau(p_j)$ to be transformed covariantly. It follows from (35)-(36) that $B_\alpha(p_j)$ under Lorentz transformations it is transformed in the same way as $b_\alpha(p)$, and $D_\tau(p_j)$ in the same way as $d_\tau(p)$.

Consider an anticommutator of the $\overline{B}_\beta(p_j)$ with $B_\alpha(p_k)$:

$$\{\overline{B}_\beta(p_j), B_\alpha(p_k)\} = \frac{1}{(\Delta^3 p_j)^2} \int_{\Delta^3 p_j} d^3 p \int_{\Delta^3 p_k} d^3 p' \, \{\overline{b}_\beta(p), b_\alpha(p')\} = \frac{1}{\Delta^3 p_j} \delta_j^k \delta_\alpha^\beta \qquad (37)$$

(there is no summation over j). Similarly,

$$\{\overline{D}_\rho(p_j), D_\tau(p_k)\} = \frac{1}{\Delta^3 p_j} \delta_j^k \delta_\tau^\rho. \qquad (38)$$

By virtue of (35) and the anticommutativity of $b_\alpha(p)$ and $b_\alpha(p')$, and also of $\overline{b}_\alpha(p)$ and $\overline{b}_\alpha(p')$

$$(B_\alpha(p_j))^2 = (B_\alpha^+(p_j))^2 = (\overline{B}_\alpha(p_j))^2 = 0. \qquad (39)$$

Similarly we obtain

$$(D_\tau(p_j))^2 = (D_\tau^+(p_j))^2 = (\overline{D}_\tau(p_j))^2 = 0. \qquad (40)$$

Relations (37)-(40) show the realization of the transition from the continuous densities of the Grassmann variables $b_\alpha(p)$, $d_\tau(p)$, $\overline{b}_\alpha(p)$ and $\overline{d}_\tau(p)$ to discrete Grassmann variables $B_\alpha(p_j)$, $D_\tau(p_j)$ and $\overline{B}_\alpha(p_j)$, $\overline{D}_\tau(p_j)$. In this case, the integrals are replaced by sums over discrete values of the impulse, and for the anticommutator of the conjugate quantities related to the same volume $\Delta^3 p_j$, the delta function is replaced by a large but finite value $\frac{1}{\Delta^3 p_j}$, and zero for those belonging to different volumes.

Consider the spinor $b_1(p_1)$ obtained from $b_1(0)$ by the boost (18), and the associated with it operator $\overline{b}_1(p_1)$:

$$b_1(p_1) = ch\frac{\varphi}{2} \frac{\partial}{\partial \theta^1(p_1)} + sh\frac{\varphi}{2} \theta^4(p_1),$$

$$\overline{b}_1(p_1) = ch\frac{\varphi}{2} \theta^1(p_1) - sh\frac{\varphi}{2} \frac{\partial}{\partial \theta^4(p_1)}.$$

Wherein

$$\{b_1(p_1), \overline{b}_1(p_1')\} = \delta(p_1 - p_1').$$

Since the direction $p_1$ is arbitrary, these relations are valid for any spatial impulses. Similarly,

$$\{\overline{B}_\beta(p_j), B_\alpha(p_k)\} = \frac{1}{(\Delta^3 p_j)^2} \int_{\Delta^3 p_j} d^3 p \int_{\Delta^3 p_k} d^3 p' \{\overline{b}_\beta(p), b_\alpha(p')\} = \frac{1}{\Delta^3 p_j} \delta^k_j \delta^\beta_\alpha.$$

Operator $b_{<\alpha>}(0)\overline{b}_{<\alpha>}(0)$ (summation over indices enclosed in triangular brackets, is not present) under Lorentz rotations transferring the spinors $b_\alpha(0)$ into $b_\alpha(p)$ goes into $b_{<\alpha>}(p)\overline{b}_{<\alpha>}(p)$. Likewise, $B_{<\alpha>}(0)\overline{B}_{<\alpha>}(0)$ goes into $B_{<\alpha>}(p_j)\overline{B}_{<\alpha>}(p_j)$.

Operators $\sqrt{\Delta^3 p_j}\, B_\alpha(p_j)$, $\sqrt{\Delta^3 p_j}\, D_\tau(p_j)$, $\sqrt{\Delta^3 p_j}\, \overline{B}_\alpha(p_j)$ and $\sqrt{\Delta^3 p_j}\, \overline{D}_\tau(p_j)$ play the role of fermionic variables, from which, by analogy with the theory of algebraic spinors [6], a fermionic vacuum can be constructed. Let

$$\Psi_{B_\alpha j} = \sqrt{\Delta^3 p_j}\, B_\alpha(p_j) \sqrt{\Delta^3 p_j}\, \overline{B}_\alpha(p_j),$$
$$\Psi_{D_\tau j} = \sqrt{\Delta^3 p_j}\, D_\tau(p_j) \sqrt{\Delta^3 p_j}\, \overline{D}_\tau(p_j),$$
$$\Psi_{V_j} = \Psi_{B_1 j}\Psi_{B_2 j}\Psi_{D_3 j}\Psi_{D_4 j}.$$

So that $(\Psi_{V_j})^2 = \Psi_{V_j}$, and the operators $\Psi_{V_j}$ can also be regarded as local fermionic vacuum in impulse space. We define the fermionic vacuum operator as

$$\Psi_V = \prod_j \Psi_{V_j},$$

where the product goes over all physically possible values of j. Obviously, $\Psi_V$ is an idempotent, but its Hermitian conjugation gives a vacuum that differs from $\Psi_V$, although equivalent to it. We define the operators of the number of particles

$$\hat{N}_\alpha(p_j) = [\overline{B}_{<\alpha>}(p_j)B_{<\alpha>}(p_j),*] = -[B_{<\alpha>}(p_j)\overline{B}_{<\alpha>}(p_j),*]$$
$$\hat{N}_\tau(p_j) = [\overline{D}_{<\tau>}(p_j)D_{<\tau>}(p_j),*] = -[D_{<\tau>}(p_j)\overline{D}_{<\tau>}(p_j),*].$$

Their effect on the vacuum, by virtue of (39) and (40), gives zero, and on the one-particle state

$$\hat{N}_\alpha(p_j)\Delta^3 p_j\, \overline{B}_\beta(p_k)\Psi_V = \Delta^3 p_j [\overline{B}_{<\alpha>}(p_j)B_{<\alpha>}(p_j), \overline{B}_\beta(p_k)]\Psi_V = \delta^k_j \delta^\beta_\alpha \overline{B}_{<\alpha>}(p_j)\Psi_V.$$

That is, the operator $\hat{N}_\alpha(p_j)\Delta^3 p_j$ corresponds to the operator of number of particles in the volume $\Delta^3 p_j$.

The energy-impulse operator can be written in the form

$$\hat{P}_\mu = \sum_j \Delta^3 p_j\, p_\mu(p_j)\, [\hat{N}_1(p_j) + \hat{N}_2(p_j) + \hat{N}_3(p_j) + \hat{N}_4(p_j),*] = \sum_j \hat{P}_\mu(p_j).$$

In a similar way, we can write the substitution in (6)-(11) of integrals by a sum:

$$\hat{\gamma}^\mu = \sum_j \hat{\gamma}^\mu(p_j).$$

The action of the operator $\hat{\gamma}^\mu \hat{P}_\mu$ on $\Psi\Psi_V$ for a spinor $\Psi$ with given energy-impulse $\hat{P}_\mu \Psi = p_\mu \Psi$ raises a problem, since

$$\hat{\gamma}^\mu \hat{P}_\mu \Psi \Psi_V = \hat{\gamma}^\mu p_\mu \Psi \Psi_V = (\hat{\gamma}^\mu p_\mu \Psi)\Psi_V + \Psi(\hat{\gamma}^\mu p_\mu \Psi_V) = m\Psi\Psi_V + \Psi(\hat{\gamma}^\mu p_\mu \Psi_V),$$

and wherein $\hat{\gamma}^\mu p_\mu \Psi_V \neq 0$. That is, the Dirac equation for the state vector is not satisfied. However, if we replace operators by discrete operators, the action of the operator $\sum_j \hat{\gamma}^\mu(p_j) p_\mu(p_j)$ on vacuum gives zero:

$$\sum_j \hat{\gamma}^\mu(p_j) p_\mu(p_j) \Psi_V = \sum_j \hat{\gamma}^\mu(p_j) p_\mu(p_j) \prod_k \Psi_{V_k} = \sum_j (\hat{\gamma}^\mu(p_j) p_\mu(p_j) \Psi_{V_j}) \prod_{k \neq j} \Psi_{V_k} =$$

$$= \sum_j (\hat{\gamma}^\mu(p_j) p_\mu(p_j) B_1(p_j)\overline{B}_1(p_j) B_2(p_j)\overline{B}_2(p_j) D_3(p_j)\overline{D}_3(p_j) D_4(p_j)\overline{D}_4(p_j)) \prod_{k \neq j} \Psi_{V_k} = \quad (41)$$

$$= \sum_j e^{iG_j}[\tilde{\gamma}^0(0)m, \frac{\partial}{\partial \theta^1(0)}\theta^1(0)\frac{\partial}{\partial \theta^2(0)}\theta^2(0)\frac{\partial}{\partial \theta^3(0)}\theta^3(0)\frac{\partial}{\partial \theta^4(0)}\theta^4(0)]e^{-iG_j} \prod_{k \neq j} \Psi_{V_k} = 0,$$

where the impulse $p_j$ is obtained from the state of rest by a transformation of the form (2), and $\hat{\gamma}^0 m$ then goes over into $\hat{\gamma}^\mu p_\mu$.

Therefore, we obtain

$$\sum_j \hat{\gamma}^\mu(p_j) \hat{P}_\mu(p_j) \Psi \Psi_V = m\Psi \Psi_V + \Psi \sum_j \hat{\gamma}^\mu(p_j) p_\mu(p_j) \Psi_V = m\Psi \Psi_V,$$

that is, the Dirac equation is satisfied. It is important to note that the fulfillment of (41) is due to the fact that $\hat{\gamma}^0(0)$ annihilates the local vacuum $\Psi_{V_0}$ corresponding to the state of rest:

$$\Psi_{V_0} = \frac{\partial}{\partial \theta^1(0)}\theta^1(0)\frac{\partial}{\partial \theta^2(0)}\theta^2(0)\frac{\partial}{\partial \theta^3(0)}\theta^3(0)\frac{\partial}{\partial \theta^4(0)}\theta^4(0),$$

$$\hat{\gamma}^0(0)\Psi_{V_0} = -[\frac{\partial}{\partial \theta^1(0)}\theta^1(0) + \frac{\partial}{\partial \theta^2(0)}\theta^2(0) + \frac{\partial}{\partial \theta^3(0)}\theta^3(0) + \frac{\partial}{\partial \theta^4(0)}\theta^4(0), \Psi_{V_0}] = 0.$$

At the same time $\hat{\gamma}^k(0)\Psi_{V_0} \neq 0$ and $\hat{\gamma}^5(0)\Psi_{V_0} \neq 0$. Therefore, the Dirac equation can be satisfied only in the case when $\hat{\gamma}^0$ has the signature +1 and corresponds to the time axis. Operators $i\hat{\gamma}^k$ and $\hat{\gamma}^5$ which has the signature +1 can not correspond to the time axis, since they do not annihilate the local vacuum $\Psi_{V_0}$, and therefore three operators $\hat{\gamma}^k$ (or two of them and $i\hat{\gamma}^5$) must correspond to the spatial axes, and $i\hat{\gamma}^5$ (or the third of the operators $\hat{\gamma}^k$) correspond to the Clifford 4-volume element.

In the Clifford algebra, gamma matrices of the same signature are equivalent, however the presence of a spinor vacuum leads to a distinction of the representation in which the gamma matrix $\gamma^0$ corresponding to the operator $\hat{\gamma}^0$ is diagonal.